\documentclass[a4paper,fleqn]{cas-sc}

\usepackage[authoryear,sort&compress,comma,square]{natbib}
\setcitestyle{numbers,square}
\usepackage{amssymb}
\usepackage{color}
\usepackage{xcolor} 
\usepackage{amsmath}
\usepackage{enumitem}
\usepackage[linesnumbered, ruled,vlined]{algorithm2e}
\usepackage{algpseudocode}
\usepackage{bm}
\usepackage{url}
\usepackage{makecell}
\usepackage{hyperref}
\usepackage{tabularx}
\usepackage{threeparttable}

\usepackage{rotating}

\def\tsc#1{\csdef{#1}{\textsc{\lowercase{#1}}\xspace}}
\tsc{WGM}
\tsc{QE}

\begin{document}
\let\WriteBookmarks\relax
\def\floatpagepagefraction{1}
\def\textpagefraction{.001}
\let\printorcid\relax

\title[mode = title]{MelodyGLM: Multi-task Pre-training for Symbolic Melody Generation}
\shorttitle{} 

\author[1]{Xinda Wu}

\author[1]{Zhijie Huang}

\author[1,2]{Kejun Zhang}
\cormark[1]
\ead{zhangkejun@zju.edu.cn}

\author[1]{Jiaxing Yu}

\author[4]{Xu Tan}

\author[1]{Tieyao Zhang}
\author[3]{Youhan Li}
\author[1]{Zihao Wang}

\author[1]{Lingyun Sun}

\cortext[cor1]{Corresponding author.}

\affiliation[1]{organization={College of Computer Science and Technology, Zhejiang University},
            city={Hangzhou},
            postcode={310063},
            country={China}}
\affiliation[2]{organization={Innovation Center of Yangtze River Delta, Zhejiang University},
            city={Jiaxing},
            postcode={314100},
            country={China}}
\affiliation[3]{organization={Department of Electrical Engineering, Columbia University},
            city={New York},
            country={United States of America}}
\affiliation[4]{organization={Microsoft Research Asia},
            city={Beijing},
            postcode={100080},
            country={China}}

\begin{abstract}
Pre-trained language models have achieved impressive results in various music understanding and generation tasks. However, existing pre-training methods for symbolic melody generation struggle to capture multi-scale, multi-dimensional structural information in note sequences, due to the domain knowledge discrepancy between text and music. Moreover, the lack of available large-scale symbolic melody datasets limits the pre-training improvement. In this paper, we propose MelodyGLM, a multi-task pre-training framework for generating melodies with long-term structure. We design the melodic n-gram and long span sampling strategies to create local and global blank infilling tasks for modeling the local and global structures in melodies. Specifically, we incorporate pitch n-grams, rhythm n-grams, and their combined n-grams into the melodic n-gram blank infilling tasks for modeling the multi-dimensional structures in melodies. To this end, we have constructed a large-scale symbolic melody dataset, MelodyNet, containing more than 0.4 million melody pieces. MelodyNet is utilized for large-scale pre-training and  domain-specific n-gram lexicon construction. Both subjective and objective evaluations demonstrate that MelodyGLM surpasses the standard and previous pre-training methods. In particular, subjective evaluations show that, on the melody continuation task, MelodyGLM gains average improvements of 0.82, 0.87, 0.78, and 0.94 in consistency, rhythmicity, structure, and overall quality, respectively. Notably, MelodyGLM nearly matches the quality of human-composed melodies on the melody inpainting task.
\end{abstract}

\begin{keywords}
Music generation \sep Melody generation \sep Melody inpainting \sep Pre-trained language models \sep N-gram \sep Transformer 
\end{keywords}

\maketitle
\section{Introduction}
Symbolic melody generation, a long-standing and popular artificial intelligence application in generative art and computational creativity, has achieved impressive results with the aid of deep learning in recent years \cite{ji2023survey, wu2020popmnet}. This field has two common scenarios: melody continuation \cite{zou2022melons,wu2020popmnet} either conditioned on past contexts or initiated from scratch, and melody inpainting \cite{Tan2022MelodyIW,wei2022music} conditioned on the surrounding context. In contrast to the development of task-specific models for different scenarios, pre-trained language models (PLMs) \cite{ding2023parameter,wang2022pre}, which have achieved remarkable performance in various natural language processing (NLP) tasks, are increasingly applied in melody generation \cite{li2023mrbert,sheng2021songmass}. The core idea is to first pre-train language models on large-scale unlabeled data to learn general representations and then fine-tune them on specific downstream tasks to effectively transfer the general knowledge to target domains. However, due to the inherent domain knowledge discrepancy between text and music, applying the successful methods of PLMs from NLP to symbolic melody generation is fraught with challenges, notably in modeling long-term structures, which include:

\begin{itemize}[itemsep=0pt, topsep=5pt, partopsep=0pt]
\item[1)] \textbf{Multi-scale modeling.} While both text and music exhibit local and global dependencies in their sequential information, their syntactic systems differ significantly in formal terms \cite{koelsch2013processing,patel2010music}. Music relies heavily on repetition to build structure and convey meaning \cite{Huang2018MusicTG}, which is essential for the music listening experience. Therefore, music structure features recurring themes, motifs, and phrases across different time scales \cite{ji2023survey,borsos2023audiolm,yu2022museformer}, a characteristic not typically observed in text data. Although many studies have attempted to generate longer music pieces with reasonable structures \cite{Huang2018MusicTG,wu2020popmnet,zou2022melons,lu2022meloform, yu2022museformer,zhang2023wuyun}, simultaneously modeling the local and global structures in music remains a significant challenge.
\item[2)] \textbf{Multi-dimensional modeling.} In musical notation, a melody is a linear sequence of musical notes, akin to a single string of text. However, music has multiple dimensions with pitch and rhythm being the most prominent ones \cite{krumhansl2000rhythm}. Not only do these dimensions have their individual structural frameworks, but they also engage in complex interactions, as evidenced by the rich literature in music cognition and psychology\cite{Prince2019ListenersPC, Prince2017SurfaceAS, Prince2014TheTH}. Therefore, in addition to multi-scale modeling, it is crucial to develop an approach that can model the structural information at several dimensions, such as pitch and rhythm, as well as their synergistic relationships.
\item[3)]\textbf{Large-scale pre-training.} To establish general composition skills, pre-training on a large-scale symbolic melody dataset \cite{zeng2021musicbert} using efficient pre-training objectives that incorporate the two fundamental melodic properties mentioned above is essential.
\end{itemize}

Recently, many pre-training frameworks based on transformer architecture have demonstrated excellent performance on both conditional and unconditional text generation tasks \cite{raffel2020exploring,bao2020unilmv2,du2022glm}. Among these frameworks, Du et al. introduce a general pre-training framework (GLM) with variable-length autoregressive blank infilling pre-training objectives. GLM can effectively model the local and global structures for the long text generation task \cite{du2022glm} and outperforms BERT \cite{kenton2019bert}, T5 \cite{raffel2020exploring}, and GPT \cite{radford2018improving} models. To further improve the performance of pre-training, there are various efficient and knowledge-enhanced masking strategies to better represent contiguous sequences of n tokens, namely n-grams, such as entities \cite{zhang2019ernie}, whole words \cite{cui2021pre}, and text spans \cite{joshi2020spanbert}. However, there is a significant difference in the structural building blocks between music and text, rendering these masking strategies less effective in capturing musical patterns in the music generation domain \cite{zeng2021musicbert}. Moreover, inadequacies in current musical structure recognition hinder researchers from adjusting these strategies for pre-training in music generation. Inspired by the analogous nature of skill acquisition in both natural language and music, we recognize that statistical language modelings, such as n-gram language modeling \cite{rohrmeier2018musical}, offer a versatile mechanism to analyze and extract useful coarse-grained linguistic information in text and musical patterns in music. Notably, n-gram language modeling is flexible to capture local patterns across multiple dimensions, either independently or jointly. To the best of our knowledge, no existing studies have effectively leveraged n-gram masking strategies to enhance pre-training for melody generation tasks. However, since both pre-training and n-gram language modeling are significantly dependent on a large-scale corpus, the scarcity of publicly available large-scale symbolic melody datasets significantly constrains the development of such methods.

In this paper, we propose MelodyGLM, a multi-task pre-training framework that allows the model to simultaneously learn multi-scale, multi-dimensional structure information for symbolic melody generation. We create local and global blank infilling tasks to jointly pre-train the model by autoregressively reconstructing short- and long-term corrupted spans for modeling the local and global structures in melodies. We carefully design the melodic n-gram and long span sampling strategies, and apply an optimal collocation ratio to achieve efficient multi-task learning. Specifically, we utilize pitch n-grams, rhythm n-grams, and combined n-grams in the melodic n-gram sampling strategy to effectively model the multi-dimensional structures in melodies. Moreover, we construct a large-scale and diverse symbolic melody dataset called MelodyNet that contains more than 0.4 million melody pieces extracted from approximately 1.6 million songs. MelodyNet is used for large-scale pre-training and domain-specific n-gram lexicon construction.

We fine-tune the pre-trained MelodyGLM on melody continuation and inpainting tasks. Our objective and subjective experiments show that MelodyGLM outperforms the standard and various advanced pre-training methods by a large margin on both tasks. In particular, subjective evaluations show that, on the melody continuation task, MelodyGLM gains average improvements of 0.82, 0.87, 0.78, and 0.94 in consistency, rhythmicity, structure, and overall quality, respectively. Remarkably, MelodyGLM nearly matches the quality of human-composed melodies on the melody inpainting task. Furthermore, ablation studies demonstrate the effectiveness of the individual components and their cooperation in MelodyGLM, as well as the benefits of the large-scale corpus MelodyNet for pre-training.

The rest of this paper is organized as follows. Section 2 briefly discusses some background material for our study. Section 3 describes the proposed MelodyGLM framework in detail. Section 4 provides a detailed description of the experimental setup. Results and method analysis are provided in Section 5. Finally, conclusions and future work are drawn in Section 6.

\section{Related works}
\subsection{Pre-training framework and objective}
 Transformer-based pre-trained models can be categorized into three main types by their network architecture: encoder-only, decoder-only, and encoder-decoder models. Encoder-only frameworks leverage a multi-layer bidirectional transformer encoder to effectively learn contextualized representations, primarily designed for understanding tasks, such as BERT \cite{kenton2019bert}. Decoder-only models employ a unidirectional transformer decoder to effectively handle longer text sequences, making it particularly suitable for generative tasks, such as GPT \cite{radford2018improving}. Conventional encoder-decoder models combine the encoder and decoder components, resulting in a stronger ability to learn and generate sequences based on specific conditions \cite{vaswani2017attention}.

The standard objective for pre-training language models (LMs) is to improve the left-to-right prediction of the next word (i.e., causal language modeling) on large-scale unlabeled corpora. Besides, denoising objectives is a popular alternative to the standard LM objective, which aims to train LMs to reconstruct original data from corrupted or noisy input \cite{liu2023pre}. One of the most widely used denoising objectives is masked language modeling (MLM). The original masking strategy, first introduced in the BERT model, consists of randomly selecting 15\% of the input tokens based on a uniform distribution. Furthermore, various sophisticated masking strategies have been proposed to introduce more diverse information and prior knowledge for LMs, such as Whole Word Masking \cite{cui2021pre}, Contiguous Spans Masking \cite{joshi2020spanbert}, PMI-Masking \cite{levine2021pmimasking}, N-gram Masking \cite{Xiao2020ERNIEGramPW,cui2020revisiting}, Entity Masking \cite{sun2019ernie}. Other noising functions like deletion, infilling, permutation, and rotation have also improved PLMs for various downstream tasks \cite{lewis2020bart}. For further details on other auxiliary training objectives, please refer to the referenced publication \cite{liu2023pre}.

Researchers have proposed more efficient frameworks and objectives to enable a pre-training framework to address diverse downstream tasks. UniLM \cite{dong2019unified} integrates the encoding and decoding steps into a single architecture with shared model parameters (known as a unified encoder-decoder architecture) and controls the range of attention for each token in the self-attentional mask matrix to flexibly tackle various language modeling tasks. T5 \cite{raffel2020exploring} formulates all text-based language tasks into a unified text-to-text format and achieves competitive performance with encoder-decoder architectures. GLM \cite{du2022glm} proposes multi-scale blank infilling training objectives with a unified encoder-decoder architecture and surpasses previous models like BERT, T5, and GPT in various NLP tasks. In this paper, we build our pre-training methods upon a similar GLM framework and extend it with several improvements to address the pre-training challenges in symbolic melody generation.

\subsection{Symbolic music generation with pre-training}
The pre-training and fine-tuning paradigm has advanced music generation but is yet to be fully adopted and customized for the domain. LakhNES \cite{donahue2019lakhnes} pre-trains a Transformer-XL \cite{dai2019TransformerXL} using standard language modeling and music data augmentation techniques on the Lakh MIDI dataset (LMD) \cite{Raffel2016LearningBasedMF} and then fine-tunes it on the NES dataset using transfer learning to further improve the performance of multi-instrumental music generation. SongMASS \cite{sheng2021songmass} leverages the MASS \cite{Song2019MASSMS} pre-training method based on the encoder-decoder framework for automatic songwriting. SongMASS lengthens the original length of masking spans from sentence level to song level for capturing longer contextual information about music repeat structure. MuseBERT \cite{wang2021musebert} pre-trains BERT on the 19.8K 2-bar piano music segments for short-term music understanding (e.g., chord analysis) and conditional music generation tasks (e.g., accompaniment generation and refinement). MuseBERT proposes a generalized relative position encoding method to better capture piano music's non-sequential, polyphonic structure. MRBERT \cite{li2023mrbert} also utilizes BERT for multi-task-based music generation tasks. MRBERT separately learns a melody's pitch and rhythm representations since music has more dimensions than text. However, MRBERT used only 452 leadsheets for pre-training and fine-tuning, which may have led to severe overfitting. Additionally, both MuseBERT and MRBERT pre-train BERT using an original random masking strategy for music generation tasks, which has been found to be inadequate for generation tasks in terms of the model's architecture \cite{du2022glm} and sampling strategies \cite{wettig2023mask}. 

Unlike existing pre-training methods for music generation, this work introduces a multi-task pre-training framework with multi-scale, multi-dimensional blank infilling for generating melodies with long-term structure. We carefully design the melodic n-gram and long span blank infilling objectives to simultaneously capture the local and global structures. Besides, the melodic n-gram objective is equipped with three types of melodic n-grams for modeling the multi-dimensional structures and interactions, including pitch n-grams, rhythm n-grams, and combined n-grams. Furthermore, we thoroughly investigate the appropriate corruption ratios for each objective. We pre-train MelodyGLM on a self-established large-scale symbolic melody dataset to cultivate a foundational understanding of intricate melodic structures.

\section{MelodyGLM pre-training framework}
In this section, we present MelodyGLM, a multi-task pre-training framework designed for melody generation. We begin by describing the extraction of melodic n-grams in Section 3.1. Next, in Section 3.2, we introduce multi-task pre-training based on auto-regressive blank infilling objectives for melody generation. Then, in Sections 3.3 and 3.4, we describe the symbolic melody representation method and provide an overview of the model architecture. Finally, Section 3.5 introduces the self-established large-scale symbolic melody dataset for pre-training.

\subsection{Melodic n-gram extraction}
\begin{figure}[t]
\centering
\includegraphics[width=0.9\textwidth]{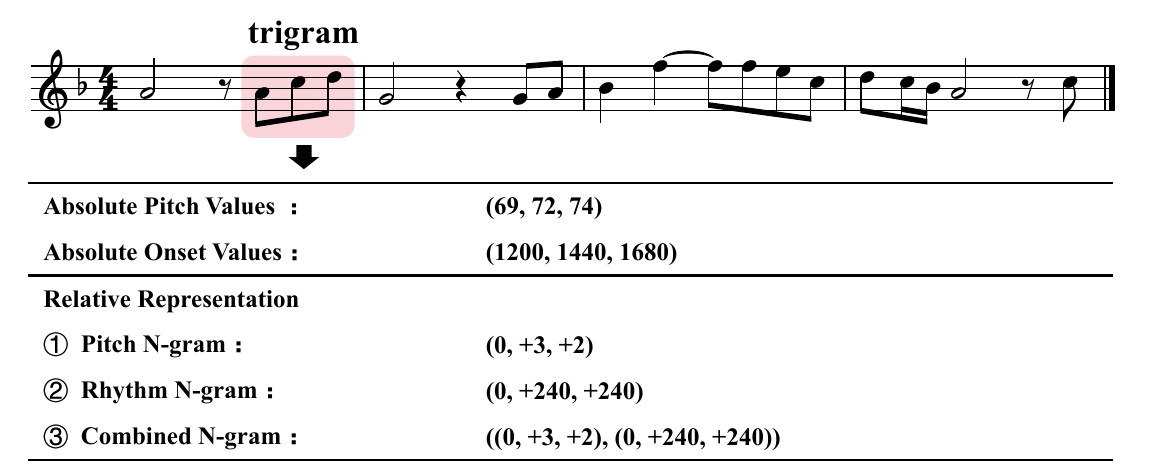}
\caption{Illustration of the relative representation for three types of melodic n-grams. A trigram case from the melody segment "Hey Jude."}
\label{fig_ngram_rep}

\centering
\includegraphics[width=0.9\textwidth]{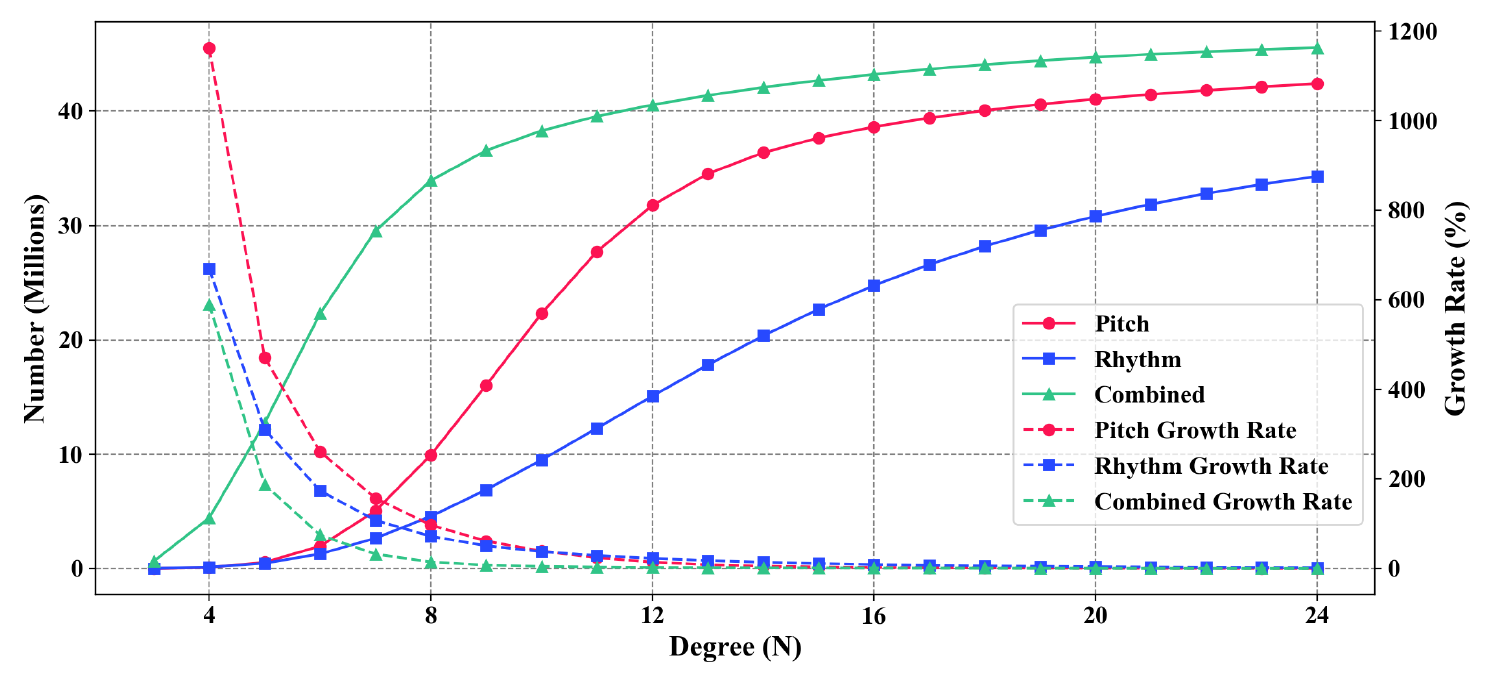}
\caption{Visualization of the number and growth trends of three types of melodic n-grams as degree 'n' ascends from 3 to 24.}
\label{fig_ngram_number}

\end{figure}

N-grams are sequences of $n$ (i.e., degree) successive items extracted from a given text or music piece. In music, n-grams can capture local melodic or harmonic patterns using statistical language modeling on musical notes or chord sequences. Therefore, musical n-grams are extensively used in Music Information Retrieval for tasks related to content understanding and generation, such as musical style and genre recognition \cite{zheng2017music},  music composition \cite{wolkowicz2008n}, and evaluation\cite{lo2006evolving, wu2023compose}.

PLMs, when employing conventional span masking techniques for music, often find it challenging to effectively capture the multi-dimensional structures and their interactions in melodic sequences \cite{zeng2021musicbert}. In this paper, we propose a novel span sampling strategy based on three types of melodic n-grams to address these challenges. These three types of melodic n-grams consist of pitch n-grams, rhythm n-grams, and combined n-grams. We adopt a relative representation method for the melodic n-grams to eliminate the influences of key and tempo in the absolute representation method. For the pitch n-gram, the pitch interval between melodic notes is utilized as the target item for extraction. Similarly, for the rhythm n-gram, the onset of melodic notes (measured in ticks) is utilized as the target item for extraction. The combined n-gram is constructed by integrating the relative pitch and rhythm n-gram from identical melodic sequences. The melodic n-grams are omitted if the rest duration between two consecutive notes exceeds one bar. As an illustration, Figure \ref{fig_ngram_rep} presents three types of melodic trigrams extracted from the "Hey Jude" melody segment.  

We analyze the degree interval of three types of melodic n-grams with degrees ranging from 3 to 24 in terms of their numbers and growth rates, as shown in Figure \ref{fig_ngram_number}. The number of three types of melodic n-grams initially shows a notable acceleration, which subsequently moderates, particularly distinguished between degrees 3 and 12. Therefore, we extract three types of melodic n-grams within the degree of 12. Furthermore, we compute the t-statistic scores for all extracted melodic n-grams, retaining only the top 25\% to form the final melodic n-gram lexicon. The higher the t-statistic score of a melodic n-gram, the more likely it is to represent a significant and recurring melodic pattern. The melodic n-gram extraction methodology is detailed in Algorithm \ref{alg_ngram}.

\IncMargin{1em}
\begin{algorithm}[t] 
    \SetKwData{Left}{left}
    \SetKwData{This}{this}
    \SetKwData{Up}{up} 
    \SetKwFunction{Union}{Union}
    \SetKwFunction{FindCompress}{FindCompress} 
    \SetKwInOut{Input}{input}
    \SetKwInOut{Output}{output}
	
	\Input{Large-scale symbolic monophonic melody corpora $M$ for pre-training} 
	\Output{Regular melodic $n$-gram lexicon $D_{N}$}
	 \BlankLine
    Dimensions $\leftarrow$ ["pitch", "rhythm", "combined"] \\
    \For{dimension $d$ in Dimensions}{
        $D_d \leftarrow \{\}$ \\
        \For{k in range(2, $n$)}{
            \emph{extract melodic n-grams in relative representation}\;
            $G_k \leftarrow []$ \Comment{Initialize a list to store each melody's $k$-grams}\\
            \For{melody $m$ in $M$}{
                $m' \leftarrow$ \textit{RelativeRepresentation}($\textit{m}$, d) \Comment{Represent each melody in relative mode}\\
                $G_{k} \leftarrow$ \textit{list(nltk.ngrams($m'$, $k$))}          \Comment{Extract melodic $k$-grams using python \textit{NLTK} toolkit}\\
            }

            \emph{select topk melodic n-grams with t-test}\;
            $P_{k} \leftarrow$ \textit{Counter($G_k$)}  \Comment{Calculate the frequency for every distinct $k$-gram}\\
            $N_k \leftarrow len(P_k)$          \Comment{Calculate the number of distinct $k$-grams in $M$}\\
            $V_{k}  \leftarrow$ \{\}           \Comment{Initialize the lexicon for melodic $k$-grams}\\
            \If{$k > 2$}{
                \For{$k$-gram $s$, frequency $p(s)$ in $P_{k}$}{
                $unigram \leftarrow$ \textit{list}(\textit{nltk.ngrams}($s$, $2$)) \\
                $p'(s) \leftarrow$  $\prod_{i=1}^{k-1}p(unigram_i)$    \Comment{Calculate the probability $p'(s)$ when the $k$-gram without statistical significance}\\
                
                $\sigma^2$ = $p(s)(1-p(s))$ \\                 
                score = $\frac{p(s)-p'(s)}{\sqrt{\sigma^2/N_{k}}}$ \Comment{Calculate the $t$-statistic score}\\
                $V_k \leftarrow V_k$ $\cup$ {($s$, score)}
                }
                $V_k \leftarrow$ \textit{topk}($V_k$, ratio)   \Comment{Select the topk melodic n-grams}\\
            }
            $D_d \leftarrow V_3 \cup ... \cup V_n$\\
        }
        $D_N \leftarrow D_d$ \Comment{Merge three types of melodic n-grams}\\
    }
    \Return $D_N$
\caption{Melodic n-grams extraction with T-test.}
\label{alg_ngram} 
\end{algorithm}
\DecMargin{1em} 

\subsection{Multi-task pre-training}
We seek to pre-train a single model capable of both unconditional and conditional melody generation tasks rather than designing multiple task-specific models. To this end, we adopt a multi-task pre-training framework with auto-regressive blank infilling objectives in MelodyGLM. Below, we detail the auto-regressive blank infilling approach \cite{du2022glm} and our proposed multi-scale, multi-dimensional auto-regressive blank infilling objectives for pre-training.  

Auto-regressive blank infilling is an effective pre-training technique for language models. It involves blanking out continuous spans of tokens from the input sequence and training the model to reconstruct these spans sequentially. Given an input sequence $\bm{x}=[x_1,\ldots,x_n]$, multiple token spans $\bm{s}=\{{s_1, \ldots, s_m}\}$ are sampled using specific strategies. Each span $s_i$ corresponds to a series of consecutive tokens $s_i = \{s_{i,1} \ldots, s_{i,l_{i}}\}$ in $\bm{x}$. To create a corrupted token sequence $x_{corrput}$, each span is replaced with a single special symbol <MASK>. The model is trained to predict the missing tokens in the spans from the corrupted text in an autoregressive manner. Formally, the probability distribution for the $i^{th}$ masked span is described as:
\begin{equation} 
 p_\theta (s_i|x_{corrupted},s_{<i}) = \prod_{j=1 }^{l_i} p_{\theta}(s_{i,j}|x_{corrupted}, s_{<i},s_{i,<j})
\end{equation}
Therefore, an auto-regressive blank infilling objective is performed to minimize the negative likelihood:
\begin{equation} 
\mathcal{L}(x, s) = -\log p_{\theta} (s|x_{corrupted}) = -\sum_{s_i\in s}\sum_{s_{i,j} \in s_i} \log p_{\theta} (s_{i,j}|x_{corrputed})
\end{equation}

We design the following pre-training objectives for modeling the multi-scale, multi-dimensional melodic structures, as well as their interactions:
\begin{itemize}[itemsep=0pt, topsep=5pt, partopsep=0pt]
\item \textbf{Melodic n-gram.} We use the Maximum Matching Algorithm \cite{Xiao2020ERNIEGramPW} to randomly sample three types of melodic n-grams in melody pieces based on the constructed melodic n-gram lexicon. The sampling ratio for each type of melodic n-grams (i.e., pitch n-grams, rhythm n-grams, and combined n-grams) accounts for 15\% of the original length. All musical note attributes of the selected melodic n-grams are masked. These objectives aim to sufficiently capture short-term melodic patterns and multi-dimensional dependencies in melodic sequences.
\item \textbf{Long span.} We uniformly sample a single long span in a melody piece, which accounts for 50\% of the original length. This objective aims to capture the long-term structure between musical elements, which is essential for the melody continuation task.
\end{itemize}
Therefore, our proposed multi-task pre-training objectives are performed by minimizing joint negative likelihood:
\begin{equation}
\mathcal{L} = \mathcal{L}_{\textit{pitch}} + \mathcal{L}_{\textit{rhythm}} + \mathcal{L}_{\textit{combined}} + \mathcal{L}_{\textit{long}}
\end{equation}
Here, $\mathcal{L}_{\textit{pitch}}$, $\mathcal{L}_{\textit{rhythm}}$, $\mathcal{L}_{\textit{combined}}$, and $\mathcal{L}_{\textit{long}}$ denote the pre-training objectives for pitch n-gram, rhythm n-gram, combined n-gram, and long span, respectively.

\subsection{Symbolic melody representation}
\begin{table}[t]
\renewcommand{\arraystretch}{1.1}
\caption{Details of our symbolic melody representation in our implementation, including token type, attribute name, value representation, and vocabulary size.}
\label{vocabulary}
\begin{tabular}{llll}
\hline
\toprule
Token Type & Attribute Name & Value Representation & Vocabulary Size \\ \hline                                                                                                       
Note    & Tempo     & \begin{tabular}[c]{@{}l@{}}Largo ($Tempo<60$), Larghetto ($Tempo \in [60, 66)$), \\Adagio ($Tempo \in [66, 76)$),  Andante ($Tempo \in [76, 108)$), \\Moderato ($Tempo \in [108, 120)$),  Allegro ($Tempo \in [120, 168)$), \\Presto ($Tempo \ge 168$) \end{tabular} & 7    \\ \cline{2-4} 
        & Bar       & Bar\_Val   ($Val \in \{x \in \mathbb{N} | 0 \le x \le 127\}$)                       & 128  \\ \cline{2-4}            
        & Position  & Pos\_Val   ($Val \in \{0, 30, 60, \ldots, 1890\}  \cup  \{0, 40, 80, \ldots, 1880\}$)       & 96   \\ \cline{2-4}       
        & Pitch     & Pitch\_Val ($Val  \in \{x \in \mathbb{N} | 0 \le x \le 127\}$)                      & 128  \\ \cline{2-4}                                     
        & Duration  & Dur\_Val ($Val  \in \{30, 60, 90, \ldots, 1920\} \cup \{40, 80, 160, 320, 640\}$)      & 96   \\ \hline                            
Special & Special   & \textless{}BOS\textgreater{}, \textless{}EOS\textgreater{}, \textless{}MASK\textgreater{}, \textless{}PAD\textgreater{}, \textless{}SEP\textgreater{}, \textless{}SEG\textgreater{}                                                                                                                    & 6    \\ 
\bottomrule
\end{tabular}
\end{table}
Inspired by the OctupleMIDI \cite{wang2021musebert} and MeMIDI \cite{zhang2023wuyun} representation methods, we design an OctupleMIDI-like symbolic melody representation for efficient pre-training and high-quality melody generation. Our token vocabulary is listed in Table \ref{vocabulary}. We classify tokens into the \textbf{\textit{Note}} and \textbf{\textit{Special}} families as follows:

For the \textbf{\textit{Note}} token type, we consolidate a set of musical attributes for a melodic note into a single compound token. Each note corresponds to a single note token that contains five musical elements: tempo, bar, position, pitch, and duration. We omit the velocity attribute from musical notes. This strategy effectively shortens the sequence length \cite{wang2021musebert} and facilitates the masking operation.

In the category of \textbf{\textit{Special}} tokens, several special delimiter symbols exist. The \textit{<BOS>} token denotes the beginning of a sequence, while the \textit{<EOS>} token denotes the end. The \textit{<MASK>} tokens are used to replace sampled spans, while the \textit{<PAD>} tokens are added to reach the maximum length. The \textit{<SEP>} token, acting as a separator, is added to the end of each sampled span. Additionally, the <SEG> token represents melodic phrase boundaries, providing valuable cues for melodic structure modeling \cite{wu2023compose, lu2022meloform}. The \textit{<SEG>} token is inserted at the detected location. The algorithm for melodic phrase boundary detection is outlined in \ref{appendix_pbda}. Like the \textbf{\textit{Note}} token, each \textbf{\textit{Special}} token contains five elements, with each element's value being the token itself.

\subsection{Model architecture}
\begin{figure}[t]
\centering
\includegraphics[width=1\textwidth]{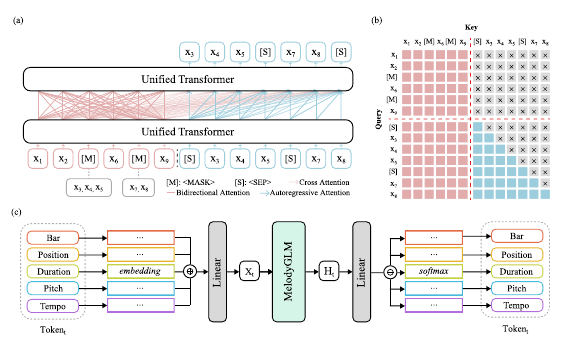}
\caption{Overview architecture of our MelodyGLM framework. (a) Detailed architecture of MelodyGLM based on auto-regressive blank infilling. (b) Self-attention mask in a unified transformer. (c) The input and output modules of our model.}
\label{fig_architecture}
\end{figure}

The overall architecture of MelodyNet is shown in Figure \ref{fig_architecture}, which adopts a single Transformer-based unified encoder-decoder framework. In Figure \ref{fig_architecture}(a), the input sequence is divided into two sections: prefix (shown in red) and suffix (shown in blue). The prefix section represents a sequence of noise-masked tokens, where a bidirectional attention mechanism is employed. This allows symbols within the prefix to attend to each other. On the other hand, the suffix section consists of one or more spans of masked tokens from the prefix section, with each span separated by the special symbol \textit{<SEP>}. Within the suffix segment, the model's attention is restricted to preceding symbols and symbols within the prefix section, facilitated by an autoregressive attention mechanism. Therefore, as illustrated in Figure \ref{fig_architecture}(b), MelodyGLM regulates the self-attention mask matrix $M_{i,j}$ to control the attention scope by modifying the attention weight $W_A$:
\begin{equation} 
M_{ij} = 
\begin{cases}
  0,& \text{  allow to attentd } \\
  -\infty,& \text{ prevent from attending } 
\end{cases}
\end{equation}
\begin{equation} 
W_A= softmax(\frac{QK^T}{\sqrt{d_k}}+M )
\end{equation}

Besides, we made several modifications to the architecture for symbolic melody generation: 
\begin{itemize}[itemsep=0pt, topsep=5pt, partopsep=0pt]
    \item[(1)] We use GeLUs \cite{hendrycks2016gaussian} instead of ReLU activation functions. 
    \item[(2)] We use $K$ separate embedding layers to encode multiple token elements in the input module of our model, as shown in the left part of Figure \ref{fig_architecture}(c). Specifically, each element $X_{t,k}$ of a token in a sequence at the $t^{th}$ time step is converted to an embedding vector via its corresponding embedding layer $Embedding(\bullet)$ with adaptive size. These embeddings are then concatenated and fed into a linear layer $FFN(\bullet)$ to obtain a single vector $X_{t,k}$ as input to MelodyGLM. Here, $K$ denotes the number of music attributes; in this case,  $K=5$.  
    \begin{equation} 
    e_{i,k} = Embedding_{k}(X_{i,k}), k=1,2,...,K,
    \end{equation}
    \\[-25pt]
    \begin{equation} 
    input_i = FFN(e_{i,1} \oplus e_{i,2}\oplus ... \oplus e_{i,k})
    \end{equation}
    In addition, we use $K$ softmax matrices to predict the different elements of the next token in the output module of our model, as shown in the right part of Figure \ref{fig_architecture}(c). Specifically, we first use a linear layer to map the hidden state of the transformer's output to a vector with the same size as the vocabulary, and then divide this vector into $K$ separate element vectors, each with their defined adaptive size.
    \item[(3)] We replace the positional encoding with the bar and position embeddings to model position information, following PopMAG \cite{ren2020popmag}.
\end{itemize}

\subsection{Pre-training corpus}
A large-scale music dataset is crucial for pre-training music generation models to capture rich knowledge and achieve superior performance. However, the largest symbolic music dataset, MMD \cite{zeng2021musicbert}, which contains over 1.5 million songs, is private due to copyright limitations and lacks a source description. Due to the scarcity of publicly available large-scale datasets for pre-training, we conduct a thorough review of relevant literature and web searches to collect symbolic music data.  
We collect approximately 1.6 million raw symbolic music data from open-source datasets and web sources and obtain 444,102 melody pieces after cleaning and deduplication.
Our web collection sources are from various well-known websites, such as 
FreeMIDI\footnote{FreeMIDI: \url{https://freemidi.org}}, 
HookTheory\footnote{HookTheory: \url{https://www.hooktheory.com}}, 
BitMIDI\footnote{BitMIDI: \url{https://bitmidi.com}}, 
MuseScore\footnote{MuseScore: \url{https://musescore.com/sheetmusic}}, 
KernScores\footnote{KernScores: \url{http://kern.ccarh.org}}, 
Kunstderfuge\footnote{Kunstderfuge: \url{https://www.kunstderfuge.com}}.
In this paper, we focus on single-track monophonic melody generation. Therefore, we exclude symbolic music data in the form of piano performances (e.g., GiantMIDI-Piano \cite{Kong2020GiantMIDIPianoAL}, MAESTRO \cite{hawthorne2018enabling}, and Pop1K7 \cite{hsiao2021cwt}). We denote our dataset as MelodyNet Dataset. We compare the sizes of various symbolic music datasets in Table \ref{database}.

\begin{table}[width=0.9\textwidth, cols=4, pos=t]
\renewcommand{\arraystretch}{1.1}
\begin{center}
\caption{Summary of public symbolic music dataset and our constructed melody dataset MelodyNet.}
\label{database}
        
\begin{tabular*}{\tblwidth}{@{\extracolsep{\fill}}llllllll}
\hline
\toprule
\textbf{Corpus}  & \textbf{Genre} & \textbf{Tracks} & \textbf{Raw}  & \textbf{Pieces} & \textbf{Hours}    & \textbf{Bars}  & \textbf{Notes} \\  \midrule
NES             & Game           & Multi-track     & 5,278         & 2,045            & 25.7             & 47,362         & 321,013         \\
POP909          & Pop            & Multi-track     & 909           & 596              & 18.9               & 22,806         & 119,448         \\
MTCL            & Folk           & Multiple        & 18,109        & 2,483            & 20.4             & 38,571         & 178,860         \\
Wikifonia       & Multiple       & Single-track    & 6,423         & 3,307            & 68.4             & 125,904        & 456,192         \\
Session         & Folk           & Single-track    & 47,519        & 8,425            & 93.3             & 172,114        & 1,150,591       \\ 
LMD             & Multiple       & Multi-track     & 176,581       & 40,664           & 1,232.0         & 1,748,229      & 7,557,123       \\
SymphonyNet     & Classical      & Multi-track     & 46,360        & 36,651           & 1,201.3          & 1,119,467      & 5,034,875       \\ \midrule
Public corpus   & Multiple       & -               & 301,179       & 94,171           & 2,660.0          & 3,273,453      & 14,181,102      \\
Web collections & Multiple       & -               & 1,296,050     & 459,175          & 12,325.0         & 17,965,194     & 79,609,047      \\ \midrule
MelodyNet       & Multiple       & Single-track    & 1,597,229     & 444,102          & 11,696.0          & 16,876,528     & 75,063,529      \\ 
\bottomrule
\end{tabular*}
\end{center}
\end{table}

\section{Experimental Setup}
\subsection{Dataset}
We use the popular Wikifonia dataset \cite{wu2020popmnet, zou2022melons, li2023mrbert,wu2020ahie} for downstream tasks in the fine-tuning stage. We extract the Wikifonia dataset from the MelodyNet dataset, which amounts to 3,028 pieces, and use the remaining part of the MelodyNet dataset for pre-training. We randomly split the Wikifonia dataset into training and test sets at a ratio of 9:1. We provide a detailed description of the data pre-processing in \ref{appendix_mdp}.

\subsection{Model configurations}
MelodyGLM uses a single standard transformer from \cite{vaswani2017attention} as our basic model structure, which consists of 4 encoder/decoder layers. The number of attention heads is 8. The model dimension and inner-layer size are set as 512 and 2048. The dropout rate is 0.1. The total number of learnable parameters is approximately 53M. The vocabulary sizes of tempo, bar, position, pitch, and duration are set as 16, 16, 128, 256, and 128.

\subsection{Pre-training and fine-tuning details}
We pre-train MelodyGLM on a single NVIDIA A100 80GB Tensor Core GPU with a batch size of 128 musical pieces over a total of 125,000 steps. The average sequence length of OctupleMIDI-like representation of a melody piece is 195 tokens. Therefore, we randomly sample segments with a length of 256 tokens for pre-training. For data augmentation, we transpose melodic pitch by a random number of semitones between $-6$ and 6. We use AdamW optimizer with $\beta_{1}=0.9$, $\beta_{2}=0.98$, $\epsilon = 10^{-6}$, and weight decay of 0.1. We employ the one-cycle learning rate policy for scheduling the learning rate. This entails a warm-up phase for the first 12,500 steps where the learning rate increases from 0 to a peak of 0.0005, after which a cosine decay is applied. We set the corruption ratios of the melodic n-gram and long span blank infilling objectives as 15\% and 50\% in the multi-task learning. The effectiveness of the corruption rations is verified in the subsubsection \ref{test_multiscale} and \ref{app_abl_multitasks}.  

We fine-tune MelodyGLM for two downstream melody generation tasks: melody continuation and inpainting. Melody continuation is to generate a coherent and musically appropriate extension for a given melody fragment or from scratch. We aim to create a 32-bar melody from scratch for the melody continuation task. Melody inpainting is to fill in the missing sections of a given melody with notes that are musically consistent with the surrounding musical context. For the melody inpainting task, we aim to fill the middle 4 bars of a given melody, guided by the preceding 6 bars and the succeeding 6 bars. Therefore, we only consider sequences that are at least 16 bars long in the MelodyNet dataset for the melody inpainting task. We set the batch sizes of the melody continuation task and the melody inpainting task as 4 and 16. For inference, we employ the temperature-controlled stochastic sampling method with top-k \cite{keskar2019ctrl}, specifically setting a temperature as 0.9 and a top-k value as 10. The learning rate is warmed up over the first 10,000 steps to a peak value of 5e-5, followed by a cosine decay until it reaches 100,000 total updates. Other parameters remain consistent with the pre-training settings.

\subsection{Baselines}
We first compare the performance of MelodyGLM without pre-training (Scratch) and with standard language models (SLM).
\begin{itemize}[itemsep=0pt, topsep=5pt, partopsep=0pt]
    \item \textbf{Scratch:} MelodyGLM is trained from scratch without pre-training, using only the fine-tuning data.
    \item \textbf{SLM:} MelodyGLM is pre-trained using a standard language model (SLM) approach \cite{liu2023pre}.
\end{itemize}
In addition, we compare our proposed multi-task learning strategy with four advanced sampling strategies:
\begin{itemize}[itemsep=0pt, topsep=5pt, partopsep=0pt]
    \item \textbf{Span:} adopts the random span sampling strategy \cite{joshi2020spanbert} for autoregressive blank infilling objectives.
    \item \textbf{Bar:} adopts the random bar sampling strategy \cite{zeng2021musicbert} for autoregressive blank infilling objectives.
    \item \textbf{Long:} adopts the uniform single long span sampling strategy \cite{du2022glm} for autoregressive blank infilling objectives.
    \item \textbf{Melodic N-gram:} adopts our proposed melodic n-gram sampling strategy for autoregressive blank infilling objectives.
\end{itemize}
Unlike music understanding tasks \cite{zeng2021musicbert}, we find that music generation tasks typically require a higher corruption ratio to capture long-term dependencies effectively, in line with the conclusions in text and music generation \cite{du2022glm, wettig2023mask, sheng2021songmass}. In this paper, we uniformly set the corruption ratio at 50\% for all the above sampling strategies to ensure a fair comparison.

\subsection{Evaluation metrics}
In this subsection, we describe the objective and subjective metrics used in this paper to evaluate the quality of AI-generated melodies for the melody continuation and inpainting tasks. These metrics have been widely adopted for the evaluation of both tasks in previous studies \cite{yang2020evaluation,ren2020popmag,yu2022museformer,wu2023compose,di2021video,Tan2022MelodyIW,shih2022theme}.

\subsubsection{Objective evaluation}
We use the following objective metrics to evaluate the similarity between AI-generated and human-composed melodies. Thus, the closer they are to the real data, the better the quality of the generated melodies. We repeat each experiment 10 times on the test set to reduce the influence of stochastic sampling.
\begin{itemize}[itemsep=0pt, topsep=5pt, partopsep=0pt]
    \item \textbf{Pitch Similarity ($\mathcal{D}_\textit{P}$):} the average overlapped area of the Pitch Class Histogram (PCH) distribution between AI-generated and human-composed melodies. PCH is a pitch-based feature to evaluate the overall tonal distribution of a piece of music. 
    \item  \textbf{Rhythm Similarity ($\mathcal{D}_\textit{R}$):} the average overlapped area of the Inter-Onset-Interval (IOI) distribution between AI-generated and human-composed melodies. IOI measures the time interval between the onsets of consecutive notes, which provides insights into the overall rhythmic patterns of a piece of music.
    \item \textbf{Structure Similarity ($\mathcal{D}_\textit{S}$): }measures the average bar-level content similarity distribution between AI-generated and human-composed melodies (i.e., similarity error \cite{yu2022museformer}).
    \item \textbf{Diversity Similarity($\mathcal{D}_\textit{D}$):} the average number similarity of distinct pitch-based n-grams between AI-generated and human-composed melodies.  $\mathcal{D}_\textit{D}$ is defined as:
    \begin{equation}
        \mathcal{D}_\textit{D} = \frac{U_n}{N_n}
    \end{equation}\\
    where $n$ represents the degree of the n-gram, $U_n$ is the number of distinct pitch-based n-grams, and $N_n$ is the total number of pitch-based n-grams. A larger $\mathcal{D}_\textit{D}$ value corresponds to greater diversity in the melody. 
    Following [4], we categorize the degrees of $n$-gram into three groups: short $\mathcal{D}_\textit{Ds}$ ($n \in [3,5]$), middle $\mathcal{D}_\textit{Dm}$ ($n \in [6,10]$), and long $\mathcal{D}_\textit{Dl}$ ($n \in [11,20]$).
\end{itemize}
Additionally, we propose two objective metrics to evaluate the overall performance of MelodyGLM on both individual and all downstream melody generation tasks.
\begin{itemize}[itemsep=0pt, topsep=5pt, partopsep=0pt]
    \item \textbf{Task Score ($TS$):} sums the rank of each metric mentioned above among compared settings as the task score. We denote the task score of the melody continuation task as $TS_c$ and the melody inpainting task as $TS_i$. In this paper, we assign equal weight to each metric and task. For instance, if a model setting ranks first across all six above metrics within a specific task compared to other model settings, it will receive a total task score of 6. Therefore, the smaller the task score, the better the model performance on this task.
    \item \textbf{Overall Rank:} the overall rank determined by the sum of task scores \cite{di2021video}. 
\end{itemize}

\subsubsection{Subjective evaluation}
Following the procedure in \cite{yu2022museformer},  we conduct a human listening test to evaluate the perceptive quality of the AI-generated and human-composed melodies since objective metrics may not be acknowledged and convincing \cite{wu2020popmnet,ren2020popmag,yang2020evaluation}. We invite 10 participants to evaluate the musical pieces with payment, where 6 evaluators can understand basic music theory. The human listening test is divided into two parts: the melody continuation listening subtest and the melody inpainting listening subtest. We apply the MelodyGLM framework in seven different settings to randomly generate or inpaint 80 melody pieces: 1) training from scratch (i.e., without pre-training); 2) pre-training with standard language modeling (SLM); 3-7) pre-training based on auto-regressive blank infilling with various advanced sampling strategies, including random span, bar, long, melodic n-gram, and multi-task. Besides, we randomly select 80 human-composed melody pieces from the test set for each subtest to serve as the ground truth. Specifically, we construct 8 groups for each participant, each containing 8 melody pieces from each setting. In experiments, subjects are asked to rate melodies on a scale from 1 (lowest) to 10 (highest) on the following four metrics:
\begin{itemize}[itemsep=0pt, topsep=5pt, partopsep=0pt]
    \item \textbf{Consistency}: Does the melody sound smooth, enjoyable, and interesting?
    \item \textbf{Rhythmicity}: Does the melody have regular beat patterns and suitable rests?
    \item \textbf{Structure}: Does the melody exhibit structural patterns, such as reasonable repetition and developed musical ideas?
    \item \textbf{Overall}: An overall score of the melody.  
\end{itemize}

In the melody inpainting listening subtest, we ask participants to focus more on the contextual melodic connections of the inpainting content. We remove the invalid ratings where z-scores exceed 2. We calculate the average scores with 95\% confidence intervals for each subjective metric.

\section{Results and Analysis}
\subsection{Main results}
\subsubsection{Objective results}
Tables \ref{tab_main_conti} and \ref{tab_main_inpaint} present the comparison of characteristic objective results for MelodyGLM across different baseline settings in the melody continuation and inpainting tasks, while Table \ref{tab_main_overall} displays the overall objective results for both tasks. Table \ref{tab_main_overall} shows that the Multi-task setting ranks first in all baseline settings and has the best task scores (lower is better) in both the melody continuation and inpainting tasks. On the one hand, as shown by the $D_S$ metric in Tables \ref{tab_main_conti} and \ref{tab_main_inpaint}, the Long setting ranks Top 3 and Top 1, excelling in modeling global structure. This demonstrates the efficacy of the long pan sampling strategy in capturing global repetitive patterns in music, in line with SongMASS's conclusions \cite{sheng2021songmass}. On the other hand, based on the $D_D$ metric in Tables \ref{tab_main_conti} and \ref{tab_main_inpaint}, the Melodic N-gram setting effectively models local structures, ranking Top 1 for melody continuation ($D_{Ds}$, $D_{Dd}$, $D_{Dl}$) and Top 2 or 3 for melody inpainting. It ranks just behind the Multi-task setting. Furthermore, as indicated by the $D_P$ and $D_R$ metrics in Tables \ref{tab_main_conti} and \ref{tab_main_inpaint}, the Multi-task setting consistently exhibits excellent modeling performance in pitch and rhythm dimensions, ranking Top 1 for melody continuation ($D_P$, $D_R$) and Top 2 for melody inpainting ($D_P$, $D_R$). The foregoing experimental findings demonstrate the capability of the Multi-task setting to model multi-scale and multi-dimensional structures in note sequences. This is attributed to the multi-task learning with the melodic n-gram and long span pre-training objectives, boosting the model's capabilities.  

Tables \ref{tab_main_conti} and \ref{tab_main_inpaint} reveal that the Multi-task setting does not surpass all other baselines in every metric. The reason is that the objective metrics we used are not independent; rather, they are interrelated. For instance, the $D_S$ and $D_D$ metrics represent melodic content's global repetitiveness and local diversity, maintaining a balanced relationship. Consequently, an increase in melodic diversity leads to reduced repetitiveness. In addition, we argue that a closer similarity, based on particular musical features between AI-generated and human-composed melodies, does not necessarily ensure an improvement in the perception of AI-generated melodies among listeners. For example, the Scratch setting ranks first in terms of $D_S$ in the melody continuation task, but has the worst rating in the listening test in terms of structure, as shown in Table \ref{subject_continuation}.

In comparison, the SLM, Span, and Bar settings manifest a notable gap in musicality during objective evaluations relative to the Multi-task setting. This suggests that these pre-training objectives might be less suited for symbolic melody generation. This experiment also demonstrates that musical bars are not efficient structural building blocks to model the long-term structure in melodies. Besides, the ascent in overall rankings for the Scratch, Multi-task (LMD), and Multi-task settings—from 8th to 4th, culminating at 1st place—in the experimental results show the beneficial role of pre-training and large-scale data in improving symbolic melody generation.

\subsubsection{Subjective results}
The subjective evaluation results of the melody continuation and inpainting listening subtests are shown in Tables \ref{subject_continuation} and \ref{subject_inpainting}. We can see that the Multi-task setting consistently surpasses other baseline settings in all objective metrics in both the melody continuation and inpainting tasks. These subjective results demonstrate that our proposed multi-task pre-training framework can effectively model melodic structures and improve musicality for symbolic melody generation. Besides, we can draw the same conclusion from the objective results that large-scale pre-training can significantly improve the quality of generated melodies. Remarkably, in the melody inpainting listening subtest, MelodyGLM nearly matches the quality of human-composed melodies. Additionally, we have found that the gap between AI-generated and human-composed melodies has become progressively less evident in the melody generation listening subtest in recent years.

\begin{table}[t]
\renewcommand{\arraystretch}{1.1}
\begin{threeparttable}
\caption{Comparison of characteristic objective results for MelodyGLM among different baseline settings in the melody continuation task (mean ± std).}
\begin{tabular*}{\tblwidth}{@{\extracolsep{\fill}}lllllll}
\hline
\toprule
\textbf{Settings} & \textbf{$\mathcal{D}_\textit{P}$(\%)$\uparrow$} & \textbf{$\mathcal{D}_\textit{R}$(\%)$\uparrow$} & \textbf{$\mathcal{D}_\textit{S}$(\%)$\downarrow$} & \textbf{$\mathcal{D}_\textit{Ds}$(\%)$\downarrow$} & \textbf{$\mathcal{D}_\textit{Dm}$(\%)$\downarrow$} & \textbf{$\mathcal{D}_\textit{Dl}$(\%)$\downarrow$} \\ \hline
Scratch   & 96.23 ± 0.90         & 95.42 ± 1.87         & \textbf{1.89 ± 0.20}     & 3.21 ± 0.08     & 7.42 ± 0.18     & 16.61 ± 0.32    \\
SLM               & 98.02 ± 0.68         & 94.61 ± 2.23         & 2.03 ± 0.21     & 3.18 ± 0.10     & 7.35 ± 0.19     & 16.28 ± 0.35    \\
Span              & 97.41 ± 0.66         & 95.40 ± 1.22         & 2.35 ± 0.20     & 3.17 ± 0.04     & 7.33 ± 0.11     & 16.31 ± 0.28    \\
Bar               & 97.31 ± 1.02         & 95.38 ± 1.14         & 2.27 ± 0.25     & 3.20 ± 0.08     & 7.43 ± 0.19     & 16.58 ± 0.39    \\
Long            & 97.93 ± 0.65         & 96.10 ± 0.68         & 2.13 ± 0.15     & 3.10 ± 0.07     & 7.14 ± 0.18     & 15.86 ± 0.40    \\
Melodic N-gram         & 97.90 ± 0.84         & 95.84 ± 1.06      & 2.45 ± 0.31    & \textbf{3.06 ± 0.04}    & \textbf{7.07 ± 0.09}     & \textbf{15.67 ± 0.18}    \\ \hline
Multi-task (LMD)   & 97.71 ± 0.74         & 96.20 ± 0.54         & 3.00 ± 0.26     & 3.17 ± 0.06     & 7.27 ± 0.13     & 16.15 ± 0.33    \\ 
Multi-task      & \textbf{98.10 ± 0.46}  & \textbf{96.34 ± 0.84  }       & 2.11 ± 0.10     & 3.10 ± 0.05     & 7.17 ± 0.13     & 15.98 ± 0.32    \\
\bottomrule
\end{tabular*}
\label{tab_main_conti}
    \begin{tablenotes}[flushleft]  
      \small  
      \item Note: the objective results of the test set for the melody continuation task are 2.82\%, 6.52\%, and 14.42\% in terms of $\mathcal{D}_\textit{Ds}$, $\mathcal{D}_\textit{Dm}$, and $\mathcal{D}_\textit{Dl}$, respectively.
    \end{tablenotes}
\end{threeparttable}
\end{table}

\begin{table}[t]
\renewcommand{\arraystretch}{1.1}
\begin{threeparttable}
\caption{Comparison of characteristic objective results for MelodyGLM among different baseline settings in the melody inpainting task (mean ± std).}

\begin{tabular*}{\tblwidth}{@{\extracolsep{\fill}}lllllll}
\hline
\toprule
\textbf{Settings} & \textbf{$\mathcal{D}_\textit{P}$(\%)$\uparrow$} & \textbf{$\mathcal{D}_\textit{R}$(\%)$\uparrow$} & \textbf{$\mathcal{D}_\textit{S}$(\%)$\downarrow$} & \textbf{$\mathcal{D}_\textit{Ds}$(\%)$\downarrow$} & \textbf{$\mathcal{D}_\textit{Dm}$(\%)$\downarrow$} & \textbf{$\mathcal{D}_\textit{Dl}$(\%)$\downarrow$} \\ \hline
Scratch   & 98.36 ± 0.23         & 96.45 ± 0.39         & 0.61 ± 0.04     & 5.81 ± 0.02     & 14.09 ± 0.07    & 34.62 ± 0.19    \\
SLM               & \textbf{98.98 ± 0.14}         & 96.52 ± 0.25         & 0.23 ± 0.04     & 5.80 ± 0.02     & 14.08 ± 0.04    & 34.52 ± 0.14    \\
Span              & 98.81 ± 0.17         & 96.71 ± 0.27         & 0.24 ± 0.04     & 5.79 ± 0.01     & 14.03 ± 0.03    & 34.42 ± 0.10    \\
Bar               & 98.80 ± 0.17         & 96.74 ± 0.27         & 0.27 ± 0.04     & 5.81 ± 0.01     & 14.08 ± 0.04    & 34.49 ± 0.12    \\
Long            & 98.85 ± 0.19         & 96.74 ± 0.22          & \textbf{0.19 ± 0.03}     & 5.79 ± 0.02     & 14.04 ± 0.05    & 34.36 ± 0.11    \\
Melodic N-gram         & 98.89 ± 0.17         & \textbf{96.84 ± 0.17}         & 0.20 ± 0.04     & 5.79 ± 0.01     & 14.03 ± 0.04    & 34.39 ± 0.12    \\ \hline
Multi-task (LMD)   & 98.89 ± 0.20         & 96.70 ± 0.19         & 0.19 ± 0.04     & 5.79 ± 0.01     & 14.04 ± 0.04    & 34.43 ± 0.13    \\
Multi-task         & 98.90 ± 0.17         & 96.79 ± 0.24         & 0.24 ± 0.04     & \textbf{5.78 ± 0.02}     & \textbf{14.00 ± 0.04}    & \textbf{34.34 ± 0.11}    \\
\bottomrule
\end{tabular*}
\label{tab_main_inpaint}
\begin{tablenotes}[flushleft]  
\small  
\item Note: the objective results of the test set for the melody inpainting task are 5.77\%, 13.98\%, and 34.22\% in terms of $\mathcal{D}_\textit{Ds}$, $\mathcal{D}_\textit{Dm}$, and $\mathcal{D}_\textit{Dl}$, respectively.
\end{tablenotes}
\end{threeparttable}
\end{table}

\begin{table}[t]
\renewcommand{\arraystretch}{1.1}
\caption{Comparison of overall objective results for MelodyGLM among different baseline settings in both the melody continuation and inpainting task.}
\begin{tabular}{lllll}
\hline
\toprule
\textbf{Settings} & \textbf{$TS_c$$\downarrow$} & \textbf{$TS_i$$\downarrow$} & \textbf{$Total \space Score$$\downarrow$} & \textbf{$Overall\space Rank$$\downarrow$} \\ \hline
Scratch              & 37                & 48                & 85             & 8             \\
SLM                  & 29                & 31                & 60             & 6             \\
Span                 & 33                & 24                & 57             & 5             \\
Bar                  & 41                & 37                & 78             & 7             \\
Long               & 17                & 21                & 38             & 3             \\
Melodic N-gram            & 18                & 15                & 33             & 2             \\ \hline
Multi-task (LMD)   & 28                & 23                & 51             & 4             \\
Multi-task           & \textbf{13  }              &\textbf{12}                & \textbf{25}             & \textbf{1 }            \\

\bottomrule
\end{tabular}
\label{tab_main_overall}
\end{table}
\begin{table}[ht]
\renewcommand{\arraystretch}{1.1}
\caption{Subjective results of the melody continuation listening subtest. The average scores are calculated with 95\% confidence intervals.}
\begin{tabular}{lllll}
\hline
\toprule
\textbf{Settings}        & \textbf{Consistency} & \textbf{Rhythmicity} & \textbf{Structure}   & \textbf{Overall}     \\ \hline
Scratch         & 6.24 ± 0.61 & 6.24 ± 0.67 & 6.27 ± 0.61 & 6.23 ± 0.64 \\
SLM             & 7.35 ± 0.37 & 7.29 ± 0.2  & 7.57 ± 0.45 & 7.43 ± 0.35 \\
Span            & 6.48 ± 0.59 & 6.48 ± 0.81 & 6.74 ± 0.74 & 6.62 ± 0.77 \\
Bar             & 6.69 ± 0.52 & 6.62 ± 0.46 & 6.73 ± 0.5  & 6.69 ± 0.46 \\
Long          & 6.95 ± 0.39 & 6.85 ± 0.53 & 7.10 ± 0.53 & 6.99 ± 0.42 \\
Melodic N-gram       & 6.74 ± 0.49 & 6.65 ± 0.53 & 6.91 ± 0.53 & 6.74 ± 0.51 \\
Multi-task      & \textbf{7.56 ± 0.69} & \textbf{7.56 ± 0.68 }& \textbf{7.67 ± 0.68} & \textbf{7.72 ± 0.61} \\ \hline
Human           & 8.13 ± 0.64 & 8.06 ± 0.45 & 8.38 ± 0.45 & 8.34 ± 0.49 \\
\bottomrule
\end{tabular}
\label{subject_continuation} 
\end{table}
\begin{table}[ht]
\renewcommand{\arraystretch}{1.1}
\caption{Subjective results of the melody inpainting listening subtest. The average scores are calculated with 95\% confidence intervals.}
\begin{tabular}{lllll}
\hline
\toprule
\textbf{Settings}        & \textbf{Consistency} & \textbf{Rhythmicity} & \textbf{Structure}   & \textbf{Overall}     \\ \hline
Scratch         & 6.41 ± 0.54 & 6.36 ± 0.41 & 6.46 ± 0.36 & 6.45 ± 0.36 \\
SLM             & 7.35 ± 0.46 & 7.23 ± 0.46 & 7.26 ± 0.63 & 7.32 ± 0.55 \\
Span            & 7.22 ± 0.34 & 7.25 ± 0.33 & 7.24 ± 0.34 & 7.28 ± 0.31 \\
Bar             & 7.35 ± 0.80 & 7.44 ± 0.65 & 7.38 ± 0.69 & 7.44 ± 0.66 \\
Long          & 7.17 ± 0.60 & 7.11 ± 0.50 & 7.22 ± 0.55 & 7.17 ± 0.46 \\
Melodic N-gram       & 7.73 ± 0.66 & 7.63 ± 0.49 & 7.60 ± 0.63 & 7.63 ± 0.52 \\
Multi-task      & \textbf{8.00 ± 0.67} & \textbf{7.80 ± 0.7}  & \textbf{7.86 ± 0.69} & \textbf{7.88 ± 0.65} \\ \hline
Human           & 8.11 ± 0.67 & 7.97 ± 0.62 & 8.09 ± 0.61 & 8.10 ± 0.65 \\
\bottomrule 
\end{tabular}
\label{subject_inpainting} 
\end{table}

\subsection{Method Analysis}
\subsubsection{Study of multi-dimensional melodic n-gram}
To verify the effectiveness of the melodic n-gram design for modeling the local and multi-dimensional structures, we compare our proposed melodic n-gram strategy (1) with different degrees (2-3) and musical dimensions (4-7)  as follows: 1) melodic 4-gram, which uses pitch 4-gram, rhythm 4-gram, and combined 4-gram; 2) melodic 8-gram, which uses pitch 8-gram, rhythm 8-gram, and combined 8-gram; 3) melodic 12-gram, which uses pitch 12-gram, rhythm 12-gram, and combined 12-gram; 4) pitch 4-gram, which only uses pitch 4-gram; 5) rhythm 4-gram, which only uses rhythm 4-gram; 6) combined 4-gram, which only uses combined 4-gram; 7) independent 4-gram, which uses pitch 4-gram and rhythm 4-gram;

In Table \ref{tab_abl_overall}, the overall analysis of the melodic n-gram is presented. We observe that the melodic 4-gram surpasses the melodic 8-gram and the melodic 12-gram in both the melody continuation and inpainting tasks. This implies that a smaller degree of melodic n-gram sampling strategy can better capture melodies' characteristics. Besides, the melodic 4-gram outperforms other melodic 4-gram variants with less dimensional pre-training objectives, ranking first in terms of $TS_c$ and $Total \space Score$ and second for $TS_i$. These rankings highlight the capacity of the melodic 4-gram setting in modeling the local and multi-dimensional structures. In this experiment, rhythm 4-gram ranks first regarding $TS_i$ in the melody inpainting task. Based on Figure \ref{fig_ngram_number}, we observe that the number of rhythm patterns is much smaller than in other dimensions. Therefore, local patterns can be easily imitated from the surrounding context based on rhythm structure in the melody inpainting task. For a detailed characteristic analysis of the melodic n-gram pre-training objectives for the melody continuation and inpainting tasks, refer to Tables \ref{tab_abl_conti} and \ref{tab_abl_inpaint} in \ref{app_3Dngram}.

\subsubsection{Study of multi-scale pre-training objectives}
\label{test_multiscale}
To verify the effectiveness of the corruption ratios for our proposed multi-task framework, we study the performance of MelodyGLM with different corruption ratios for each pre-training objective and their optimal collocation. Utilizing the greedy search method, we first identified the optimal corruption ratio for the long span pre-training objective, followed by the optimal ratio for the melodic n-gram pre-training objectives. For the long span pre-training objective, we adopt corruption ratios of 50\%, 60\%, and 80\%. Ratios below 50\% are not considered, as the main goal of this objective is to capture the global structures. Empirically, a 50\% ratio is a commonly used parameter setting in long text and music generation as supported by studies like \cite{Song2019MASSMS, sheng2021songmass, du2022glm}. For the melodic n-gram pre-training objective, we choose corruption ratios of 10\%, 15\%, 20\%, 25\%, 30\%, 40\%, and 50\%, granularity from fine to coarse. Here, the melodic n-gram primarily focuses on the local structures; thus, we limited its corruption ratio to 50\% or less.

Table \ref{abl_single_overall} presents the overall objective analysis of corruption ratios for the long span pre-training objective. It reveals that the 50\% corruption ratio takes the lead, surpassing both the 60\% and 80\% ratios in terms of $TS_c$ and $TS_i$. These findings suggest that a larger corruption ratio in this context might compromise the model's efficacy in melody continuation and inpainting tasks, although it benefits the global structure modeling in the melody continuation task.

Table \ref{abl_multitask_overall} provides the overall objective analysis of cooperative corruption ratios in the Multi-task setting. It reveals that, when paired with a 50\% corruption ratio for the long span pre-training objective, the 15\% corruption ratio of the melodic n-gram pre-training objectives yields the best performance in terms of both $TS_c$ and $TS_i$. For a detailed characteristic analysis of the multi-task design with different corruption ratios in both melody continuation and inpainting tasks, refer to Tables \ref{app_long_conti}, \ref{app_long_inpaint}, \ref{app_multitask_conti}, and \ref{app_multitask_inpaint} in \ref{app_abl_multitasks}.

\begin{table}[ht]
\renewcommand{\arraystretch}{1.1}
\caption{Overall analysis of the melodic n-gram design for both melody continuation and inpainting tasks in terms of degree and dimension (mean ± std).}
\begin{tabular}{lllll}
\hline
\toprule
\textbf{Settings} & \textbf{$TS_c$$\downarrow$} & \textbf{$TS_i$$\downarrow$} & \textbf{$Total \space Score$$\downarrow$} & \textbf{$Overall\space Rank$$\downarrow$} \\ \hline
Melodic 4-gram          & \textbf{11}     & \textbf{17}     & \textbf{28}    & \textbf{1}    \\
Melodic 8-gram       & 27     & 20     & 47    & 4    \\
Melodic 12-gram      & 30     & 35     & 65    & 7    \\ \hline
Pitch 4-gram       & 32     & 30     & 62    & 6    \\
Rhythm 4-gram      & 18     & \textbf{13}     & 31    & 2    \\
Combined 4-gram    & 21     & 18     & 39    & 3    \\
Independent 4-gram & 29     & 32     & 61    & 5    \\ 
\bottomrule
\end{tabular}
\label{tab_abl_overall}
\end{table}

\begin{table}[ht]
\renewcommand{\arraystretch}{1.1}
\caption{Overall analysis of the long span design with different corruption ratios in both the melody continuation and inpainting tasks.}
\begin{tabular}{llllll}
\hline
\toprule
\textbf{Settings} &\textbf{C.R.} & \textbf{$TS_c$$\downarrow$} & \textbf{$TS_i$$\downarrow$} & \textbf{$Total \space Score$$\downarrow$} & \textbf{$Overall\space Rank$$\downarrow$} \\ \hline
Long &50\% & \textbf{8}        & \textbf{8}        & \textbf{16}    & \textbf{1 }   \\
Long &60\% & 12       & 18       & 30    & 3    \\
Long &80\% & 16       & 9        & 25    & 2    \\
\bottomrule
\end{tabular}
\label{abl_single_overall}
\end{table}

\begin{table}[]
\renewcommand{\arraystretch}{1.1}
\caption{Overall analysis of the multi-task design with different corruption ratios in both the melody continuation and inpainting tasks. While the corruption ratio for the long span is fixed at 50\%, the ratio for the melodic n-gram varies.}
\begin{tabular}{llllll}
\hline
\toprule
\textbf{Settings} &\textbf{C.R.} & \textbf{$TS_c$$\downarrow$} & \textbf{$TS_i$$\downarrow$} & \textbf{$Total \space Score$$\downarrow$} & \textbf{$Overall\space Rank$$\downarrow$} \\ \hline
Multi-task & 50\%/10\% & 24       & 30       & 54    & 5    \\
Multi-task & \textbf{50\%/15\%} & \textbf{12}       & \textbf{16}       & \textbf{28}    & \textbf{1}    \\
Multi-task & 50\%/20\% & 18       & 19       & 37    & 2    \\
Multi-task & 50\%/25\% & 27       & 18       & 45    & 3    \\
Multi-task & 50\%/30\% & 31       & 31       & 62    & 7    \\
Multi-task & 50\%/40\% & 26       & 35       & 61    & 6    \\
Multi-task & 50\%/50\% & 30       & 17       & 47    & 4    \\
\bottomrule
\end{tabular}
\label{abl_multitask_overall}
\end{table}

\subsection{Case Study}
\begin{figure}[ht]
\centering
\includegraphics[width=0.9\textwidth]{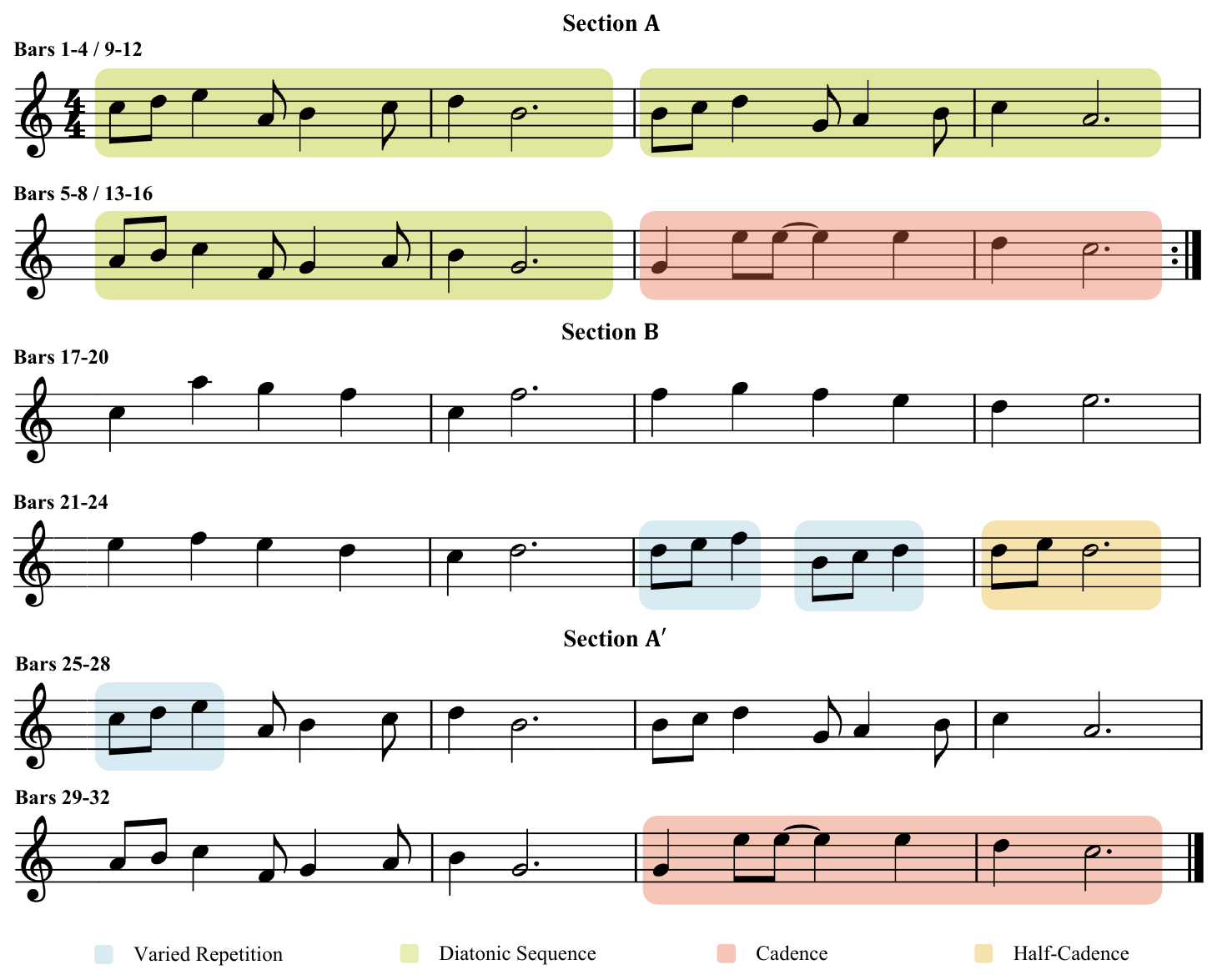}
\caption{A generated melody case from scratch by MelodyGLM.}
\label{fig_CS1}
\end{figure}

\begin{figure}[ht]
\centering
\includegraphics[width=0.9\textwidth]{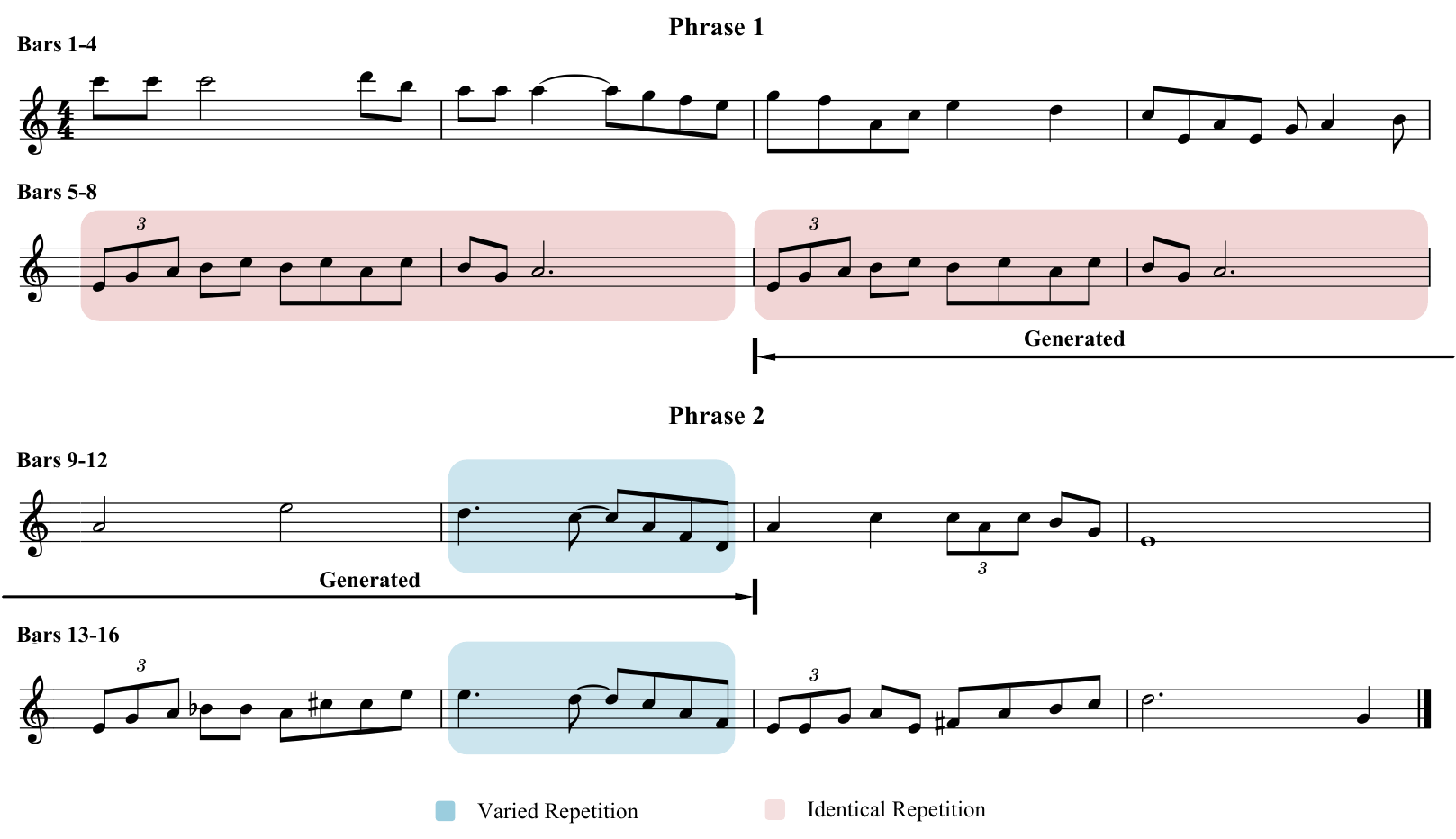}
\caption{An inpainted melody case by MelodyGLM.}
\label{fig_CS2}
\end{figure}

Figure \ref{fig_CS1} presents a 32-bar melody generated from scratch by MelodyGLM. We observe that the overall structure of this melodic segment is clearly presented in the ABA' form, corresponding respectively to A: Bars 1-16, B: Bars 17-24, and A': Bars 25-32. Section A consists of two identical phrases, each containing four 2-bar motifs. In section A, the first three motifs of each phrase (Bars 1-6 or Bars 9-14) follow the classical form of diatonic sequence \footnote{A diatonic sequence is a musical construction where an interval pattern is repeated at a higher or lower pitch, maintaining interval sizes but possibly altering interval qualities to fit the diatonic system.}, while Bars 7-8 and Bars 15-16 display a typical cadence. Section B contrasts sharply with section A, enriching the melody. Rhythmically, section A mainly exhibits syncopation, while most notes in section B are precisely on the downbeat. Notably, Bars 23-24 adopt specific elements from section A, further strengthening the thematic connection between sections B and A. It also acts as a half-cadence, seamlessly leading to section A'. Section A' is a recapitulation of section A. It is worth emphasizing that without explicit control over the formal structure of the generated melody, the model has automatically learned the formal structure knowledge from large-scale unlabeled data. This case study fully demonstrates the effectiveness of our model in capturing both short- and long-term melodic structures and enhancing the musicality of the generated melody.

Figure \ref{fig_CS2} presents a 16-bar melody inpainted by MelodyGLM, which requires the model to fill Bars 7-10 based on Bars 1-6 and Bars 11-16 content. From a global perspective, this melody segment comprises two musical phrases within these 16 measures: Bars 1-8 (Phrase 1) and Bars 9-16 (Phrase 2). The infilling contents appear in the last two bars of Phrase 1 and the first two bars of Phrase 2. The generated Bars 7-8 represent a reasonable repetition of Bars 5-6 in Phrase 1, thereby improving the structural stability of Phrase 1. Meanwhile, the agogic note at the end of Bar 8 gives a stronger sense of closure to the entire phrase. The generated Bars 9-10 introduce innovative material in terms of rhythm and pitch for Phrase 2, enriching the melodic content. Rhythmically, the pattern in Bar 9 is unique within the entire piece. Regarding pitch, the two notes, A and E, in Bar 9 form a perfect fifth interval, outlining the chord of the A minor tonic, thus continuing the tonal characteristics of the previous section. Furthermore, Bar 10 establishes a varied repetition relationship with Bar 14. In conclusion, our analysis reveals that this model exhibits proficiency in learning structural information from a musical context. In addition, it succeeds in fostering musical creativity within appropriate boundaries.

\section{Discussion}
To address the challenges of the multi-scale, multi-dimensional structure modeling in symbolic melody generation with pre-training, we introduce MelodyGLM with multi-task learning for generating melodies with long-term structure. MelodyGLM leverages local and global blank infilling tasks tailored by the melodic n-gram and long span sampling strategies to capture the local and global structures in note sequences. Simultaneously, MelodyGLM improves the melodic n-gram blank infilling pre-training with multi-dimensional structure modeling on pitch, rhythm, and their combination. Indispensably, we construct MelodyNet, a large-scale symbolic melody dataset for domain-specific n-gram lexicon construction and pre-training improvement, which contains 0.4 million diverse symbolic melodies. Both subjective and objective evaluations demonstrate that MelodyGLM can create high-quality melodies with well-formed structures and musicality on the melody continuation and inpainting tasks. Ablation studies verify the effectiveness of the component design in MelodyGLM for modeling multi-scale, multi-dimensional structures in melodies, as well as the benefit of large-scale pre-training.

Future research stemming from our study could either refine the existing framework or expand upon it. Firstly, we emphasize the achievement of controllable music generation to enhance interactivity and real-world application. Second, given our focus on symbolic melody generation, we will consider enabling the model to cater to a broader spectrum of symbolic music generation and understanding tasks. Lastly, we strive for a symphony between generated outputs and human aesthetics by incorporating advanced methodologies such as reinforcement learning from human feedback.

\section*{Declaration of competing interest}
The authors declare that they have no known competing financial interests or personal relationships that could have appeared to influence the work reported in this paper.

\section*{Data availability}
We provide download links or websites for the collected datasets in MelodyNet and AI-generated melody samples by MelodyGLM at \url{https://nextlab-zju.github.io/melodyglm/}. We release the detailed data processing procedure code at \url{https://github.com/NEXTLab-ZJU/MelodyGLM}.

\section*{Acknowledgement}
This work was supported by the National Natural Science Foundation of China (No.62272409); the Key R\&D Program of Zhejiang Province (No.2022C03126); the Project of Key Laboratory of Intelligent Processing Technology for Digital Music (Zhejiang Conservatory of Music); and the Ministry of Culture and Tourism (No.2022DMKLB001).









\appendix
\renewcommand{\thesection}{\appendixname~\Alph{section}}
\renewcommand{\thetable}{\Alph{section}.\arabic{table}}

\section{Melodic phrase boundary detection algorithm}
\label{appendix_pbda}
The melodic phrase boundary is heuristically determined by 1) the longest note in each measure that exceeds a quarter note in duration, and 2) notes that end with a rest lasting at least as long as an eighth rest \cite{lu2022meloform}. The detailed melodic phrase boundary detection algorithm is shown in Algorithm \ref{alg_rest}.

\label{alg_pbda}
\IncMargin{1em}
\begin{algorithm} \SetKwData{Left}{left}\SetKwData{This}{this}\SetKwData{Up}{up} \SetKwFunction{Union}{Union}\SetKwFunction{FindCompress}{FindCompress} \SetKwInOut{Input}{input}\SetKwInOut{Output}{output}
	
	\Input{a piece of monophonic melody's notes $N = \{n_{1},n_{2}, \ldots, n_{x}\}$} 
	\Output{a piece of monophonic melody's phrase ending notes $U = \{r_{1}, r_{2}, \ldots, r_{n}\}$}
	 \BlankLine 
	 
      $L \leftarrow  LongNotes(N)$  \Comment{Extract the long notes exceeding one beat from each measure in M}\\
      $P \leftarrow RestNotes(N)$  \Comment{Extract the first note of pairs with an interval $\geq$ the 8\-th note} \\
      $U \leftarrow L | P = \{u_{1}, u_{2}, \ldots, u_{m}\} $ \Comment{Calculate the union of two lists}\\
      
     \emph{process the start and end notes when they are in $U$}\;
     \If{$n_{1}$ in U and $u_{1}$.duratioon - $u_{2}$.duration < 240}{remove $u_{1}$ from $U$}
     \If{$n_{x}$ in $U$}{remove $n_{x}$ from $U$}
    
      \emph{remove one of continuous phrase ending notes}\;
      $i \leftarrow  1$ \\
      \While{$i$ < len(U)}{
        \If{U[i-1] and U[i] are adjacent in N}{
            \If{U[i-1].duration - U[i].duration > 240}{
                remove $U[i]$
            }
            \lElse{
            remove $U[i-1]$
            }
        }
        $i \leftarrow i + 1$
    }

    \Return $U$
\caption{Recognizing melodic phrase boundaries in a monophonic melody.}
\label{alg_rest} 
\end{algorithm}
\DecMargin{1em} 

\section{Data pre-processing}
\label{appendix_mdp}
The quality of the pre-training dataset significantly influences the performance of large language models. After collecting a large amount of symbolic music data, we employ a preprocessing pipeline to eliminate noisy, irrelevant, and redundant data, aiming to construct a high-quality melody dataset.

We employ the MuseScore\footnote{MuseScore: \url{https://musescore.org/}} software to transform several symbolic music data formats into MIDI. Melody tracks in a 4/4 time signature are extracted using MIDI Miner \cite{guo2019midi}. To address timing imprecision, we implement a self-adaptive mixed precision quantization approach (64th notes and triplets) \cite{zhang2023wuyun}. Moreover, we formulate a heuristic rule set for eliminating low-quality data, detailed in Table \ref{app_hfr}. To mitigate issues of diminishing generative diversity and potential dataset contamination, we apply both internal and external de-duplication on our pre-training dataset by the hash value of pitch interval sequences. This de-duplication approach is motivated by our observation that some melody samples are mere transpositions of others, shifted by a set number of semitones.

\begin{table}[t]
\renewcommand{\arraystretch}{1.1}
    \begin{center}
    \caption{Heuristic filtering rules for data pre-processing.}
    \label{app_hfr}
\begin{tabular}{ll}
\hline
\textbf{Type}   & \textbf{Rules}  \\ \hline                                                                                                         
\multirow{5}{*}{Musical Features} & \multirow{5}{*}{\begin{tabular}[c]{@{}l@{}}1) The number of notes in the sample is at least 32.\\ 2) The number of non-empty bars of the sample is at least 8 bars, accounting for more than 70\% \\ of the total number of bars.\\ 3) The number of consecutive identical pitches of the sample does not exceed 10.\\ 4) The number of pitch classes in the sample is greater than 5.\\ \end{tabular}} \\
& \\    & \\     & \\       &  \\ \hline                                                                                                                                                                                  
\end{tabular}
\end{center}
\end{table}

\section{Characteristic analysis of melodic n-gram with multiple dimensions}
\label{app_3Dngram}
The characteristic analysis of the melodic n-gram design for the melody continuation and inpainting tasks are detailed in Tables \ref{tab_abl_conti} and \ref{tab_abl_inpaint}.

\section{Characteristic analysis of multi-task with different scale}
\label{app_abl_multitasks}
The characteristic analysis of the corruption ratios for the multi-task pre-training objectives in the melody continuation and inpainting tasks are detailed in Tables \ref{app_long_conti}, \ref{app_long_inpaint}, \ref{app_multitask_conti}, and \ref{app_multitask_inpaint}.

\begin{table}[ht]
\renewcommand{\arraystretch}{1.1}
\caption{Characteristic analysis of the melodic n-gram design for the melody continuation task in terms of degree and dimension (mean ± std).}
\begin{tabular}{lllllll}
\hline
\toprule
\textbf{Settings} & \textbf{$\mathcal{D}_\textit{P}$(\%)$\uparrow$} & \textbf{$\mathcal{D}_\textit{R}$(\%)$\uparrow$} & \textbf{$\mathcal{D}_\textit{S}$(\%)$\downarrow$} & \textbf{$\mathcal{D}_\textit{Ds}$(\%)$\downarrow$} & \textbf{$\mathcal{D}_\textit{Dm}$(\%)$\downarrow$} & \textbf{$\mathcal{D}_\textit{Dl}$(\%)$\downarrow$} \\ \hline
Melodic 4-gram          & 97.90 ± 0.84 & \textbf{95.84 ± 1.06} & 2.45 ± 0.31 & \textbf{3.06 ± 0.04} & \textbf{7.07 ± 0.09} & \textbf{15.67 ± 0.18} \\
Melodic 8-gram       & 97.69 ± 1.03 & 95.31 ± 1.56 & 2.43 ± 0.12 & 3.17 ± 0.07 & 7.31 ± 0.15 & 16.29 ± 0.34 \\
Melodic 12-gram      & 97.80 ± 0.88 & 93.80 ± 1.13 & 2.06 ± 0.19 & 3.23 ± 0.08 & 7.53 ± 0.18 & 16.70 ± 0.33 \\ \hline
Pitch 4-gram       & 97.87 ± 0.61 & 91.61 ± 3.33 & \textbf{1.94 ± 0.17} & 3.35 ± 0.14 & 7.77 ± 0.26 & 17.06 ± 0.46 \\
Rhythm 4-gram      & \textbf{98.16 ± 0.35} & 94.22 ± 4.16 & 2.17 ± 0.14 & 3.16 ± 0.13 & 7.22 ± 0.12 & 16.07 ± 0.31 \\
Combined 4-gram    & 97.05 ± 1.12 & 95.38 ± 1.48 & 2.53 ± 0.21 & 3.12 ± 0.10 & 7.22 ± 0.23 & 16.01 ± 0.44 \\
Independent 4-gram & 96.96 ± 0.86 & 95.28 ± 1.03 & 2.71 ± 0.28 & 3.15 ± 0.06 & 7.26 ± 0.15 & 16.10 ± 0.27 \\ 
\bottomrule
\end{tabular}
\label{tab_abl_conti}
\end{table}
\begin{table}[ht]
\renewcommand{\arraystretch}{1.1}
\caption{Characteristic analysis of the melodic n-gram design for the melody inpainting task in terms of degree and dimension (mean ± std).}
\begin{tabular}{lllllll}
\hline
\toprule
\textbf{Settings} & \textbf{$\mathcal{D}_\textit{P}$(\%)$\uparrow$} & \textbf{$\mathcal{D}_\textit{R}$(\%)$\uparrow$} & \textbf{$\mathcal{D}_\textit{S}$(\%)$\downarrow$} & \textbf{$\mathcal{D}_\textit{Ds}$(\%)$\downarrow$} & \textbf{$\mathcal{D}_\textit{Dm}$(\%)$\downarrow$} & \textbf{$\mathcal{D}_\textit{Dl}$(\%)$\downarrow$} \\ \hline
Melodic 4-gram          & \textbf{98.89 ± 0.17} & 96.84 ± 0.17 & 0.20 ± 0.04 & 5.79 ± 0.01 & 14.03 ± 0.04 & 34.39 ± 0.12 \\
Melodic 8-gram          & 98.74 ± 0.28 & 96.89 ± 0.24 & \textbf{0.19 ± 0.03} & 5.79 ± 0.02 & 14.02 ± 0.04 & 34.32 ± 0.14 \\
Melodic 12-gram         & \textbf{98.89 ± 0.17} & 96.72 ± 0.23 & 0.24 ± 0.04 & 5.81 ± 0.02 & 14.09 ± 0.06 & 34.55 ± 0.18 \\ \hline
Pitch 4-gram       & 98.86 ± 0.18 & 96.81 ± 0.20 & 0.21 ± 0.05 & 5.80 ± 0.02 & 14.08 ± 0.05 & 34.48 ± 0.14 \\
Rhythm 4-gram      & 98.84 ± 0.21 & \textbf{96.98 ± 0.11} & 0.21 ± 0.04 & \textbf{5.77 ± 0.02} & \textbf{13.99 ± 0.04} & \textbf{34.26 ± 0.11} \\
Combined 4-gram    & 98.83 ± 0.22 & 96.83 ± 0.23 & 0.20 ± 0.04 & 5.78 ± 0.01 & 14.01 ± 0.04 & 34.31 ± 0.12 \\
Independent 4-gram & 98.86 ± 0.20 & 96.79 ± 0.17 & 0.25 ± 0.04 & 5.80 ± 0.02 & 14.05 ± 0.04 & 34.47 ± 0.12 \\ 
\bottomrule
\end{tabular}
\label{tab_abl_inpaint}
\end{table}
\begin{table}[ht]
\renewcommand{\arraystretch}{1.1}
\caption{Characteristic analysis of the long span design with different corruption ratios in the melody continuation tasks (mean ± std).}
\begin{tabular}{llllllll}
\hline
\toprule
\textbf{Settings} &\textbf{C.R.} &\textbf{$\mathcal{D}_\textit{P}$(\%)$\uparrow$} & \textbf{$\mathcal{D}_\textit{R}$(\%)$\uparrow$} & \textbf{$\mathcal{D}_\textit{S}$(\%)$\downarrow$} & \textbf{$\mathcal{D}_\textit{Ds}$(\%)$\downarrow$} & \textbf{$\mathcal{D}_\textit{Dm}$(\%)$\downarrow$} & \textbf{$\mathcal{D}_\textit{Dl}$(\%)$\downarrow$} \\ \hline
Long & 50\% & \textbf{97.93 ± 0.65} & \textbf{96.10 ± 0.68} & 2.13 ± 0.15 & \textbf{3.10 ± 0.07} & \textbf{7.14 ± 0.18} & \textbf{15.86 ± 0.40} \\
Long & 60\% & 97.87 ± 0.57 & 96.07 ± 0.61 & 2.11 ± 0.16 & 3.13 ± 0.06 & 7.23 ± 0.16 & 16.12 ± 0.36 \\
Long & 80\% & 97.82 ± 0.77 & 95.91 ± 1.07 & \textbf{2.10 ± 0.19} & 3.25 ± 0.06 & 7.49 ± 0.14 & 16.69 ± 0.36 \\
\bottomrule
\end{tabular}
\label{app_long_conti}
\end{table}
\begin{table}[ht]
\renewcommand{\arraystretch}{1.1}
\caption{Characteristic analysis of the long span design with different corruption ratios in the melody inpainting tasks (mean ± std).}
\begin{tabular}{llllllll}
\hline
\textbf{Settings} &\textbf{C.R.} & \textbf{$\mathcal{D}_\textit{P}$(\%)$\uparrow$} & \textbf{$\mathcal{D}_\textit{R}$(\%)$\uparrow$} & \textbf{$\mathcal{D}_\textit{S}$(\%)$\downarrow$} & \textbf{$\mathcal{D}_\textit{Ds}$(\%)$\downarrow$} & \textbf{$\mathcal{D}_\textit{Dm}$(\%)$\downarrow$} & \textbf{$\mathcal{D}_\textit{Dl}$(\%)$\downarrow$} \\ \hline
Long & 50\% & 98.85 ± 0.19 & \textbf{96.74 ± 0.22} & 0.19 ± 0.03 & \textbf{5.79 ± 0.02} & \textbf{14.04 ± 0.05} & \textbf{34.36 ± 0.11} \\
Long & 60\% & 98.83 ± 0.15 & 96.51 ± 0.19 & 0.24 ± 0.05 & 5.81 ± 0.02 & 14.10 ± 0.05 & 34.56 ± 0.12 \\
Long & 80\% & \textbf{98.86 ± 0.16} & 96.66 ± 0.21 & \textbf{0.18 ± 0.05} & \textbf{5.79 ± 0.02} & 14.05 ± 0.06 & 34.45 ± 0.16 \\
\bottomrule
\end{tabular}
\label{app_long_inpaint}
\end{table}
\begin{table}[ht]
\renewcommand{\arraystretch}{1.1}
\caption{Characteristic analysis of the multi-task design with different corruption ratios in the melody continuation tasks (mean ± std). While the corruption ratio for the long span is fixed at 50\%, the corruption ratio for the melodic n-gram varies.}

\setlength{\tabcolsep}{2mm}{
\begin{tabular*}{\tblwidth}{@{\extracolsep{\fill}}llllllll}
\hline
\toprule
 \textbf{Settings} & \textbf{C.R.} & \textbf{$\mathcal{D}_\textit{P}$(\%)$\uparrow$} & \textbf{$\mathcal{D}_\textit{R}$(\%)$\uparrow$} & \textbf{$\mathcal{D}_\textit{S}$(\%)$\downarrow$} & \textbf{$\mathcal{D}_\textit{Ds}$(\%)$\downarrow$} & \textbf{$\mathcal{D}_\textit{Dm}$(\%)$\downarrow$} & \textbf{$\mathcal{D}_\textit{Dl}$(\%)$\downarrow$} \\ \hline
 Multi-task &50\%/10\% & 96.91 ± 0.87 & 95.59 ± 0.93 & 2.62 ± 0.26 & 3.11 ± 0.06 & 7.20 ± 0.15 & 15.96 ± 0.35 \\
Multi-task &50\%/15\% & 98.10 ± 0.46 & \textbf{96.34 ± 0.84} & \textbf{2.11 ± 0.10} & 3.10 ± 0.05 & 7.17 ± 0.13 & 15.98 ± 0.32 \\
 Multi-task &50\%/20\% & 96.99 ± 0.83 & 95.16 ± 0.55 & 2.70 ± 0.22 & \textbf{3.03 ± 0.06} & \textbf{7.02 ± 0.14} & \textbf{15.53 ± 0.30} \\
 Multi-task &50\%/25\% & 97.93 ± 0.65 & 94.57 ± 1.16 & 2.47 ± 0.24 & 3.10 ± 0.08 & 7.21 ± 0.22 & 16.00 ± 0.47 \\
 Multi-task &50\%/30\% & \textbf{98.24 ± 0.33} & 92.84 ± 1.97 & 2.22 ± 0.15 & 3.26 ± 0.11 & 7.62 ± 0.25 & 16.87 ± 0.30 \\
 Multi-task &50\%/40\% & 97.86 ± 0.78 & 95.07 ± 1.21 & 2.88 ± 0.13 & 3.11 ± 0.08 & 7.18 ± 0.23 & 15.95 ± 0.44 \\
 Multi-task &50\%/50\% & 97.58 ± 0.88 & 95.20 ± 2.26 & 2.75 ± 0.18 & 3.12 ± 0.10 & 7.20 ± 0.21 & 15.99 ± 0.37 \\
\bottomrule
\end{tabular*}}
\label{app_multitask_conti}
\end{table}

\begin{table}[ht]
\renewcommand{\arraystretch}{1.1}
\caption{Characteristic analysis of the multi-task design with different corruption ratios in the melody inpainting tasks (mean ± std). While the corruption ratio for the long span is fixed at 50\%, the ratio for the melodic n-gram varies.}
\setlength{\tabcolsep}{2mm}{
\begin{tabular*}{\tblwidth}{@{\extracolsep{\fill}}llllllll}
\hline
\toprule
\textbf{Settings} & \textbf{C.R.} & \textbf{$\mathcal{D}_\textit{P}$(\%)$\uparrow$} & \textbf{$\mathcal{D}_\textit{R}$(\%)$\uparrow$} & \textbf{$\mathcal{D}_\textit{S}$(\%)$\downarrow$} & \textbf{$\mathcal{D}_\textit{Ds}$(\%)$\downarrow$} & \textbf{$\mathcal{D}_\textit{Dm}$(\%)$\downarrow$} & \textbf{$\mathcal{D}_\textit{Dl}$(\%)$\downarrow$} \\ \hline
Multi-task & 50\%/10\% & 98.83 ± 0.17 & 96.70 ± 0.19 & 0.27 ± 0.03 & 5.79 ± 0.02 & 14.05 ± 0.05 & 34.43 ± 0.13 \\
Multi-task &50\%/15\% & 98.90 ± 0.17 & 96.79 ± 0.24 & 0.24 ± 0.04 & 5.78 ± 0.02 & \textbf{14.00 ± 0.04} & \textbf{34.34 ± 0.11} \\
Multi-task &50\%/20\% & 98.82 ± 0.18 & \textbf{96.99 ± 0.18} & 0.25 ± 0.05 & \textbf{5.78 ± 0.01} & 14.02 ± 0.04 & 34.35 ± 0.11 \\
Multi-task &50\%/25\% & 98.79 ± 0.23 & 96.89 ± 0.15 & 0.23 ± 0.04 & \textbf{5.78 ± 0.01} & 14.01 ± 0.04 & 34.34 ± 0.12 \\
Multi-task &50\%/30\% & 98.87 ± 0.23 & 96.65 ± 0.10 & 0.22 ± 0.05 & 5.80 ± 0.01 & 14.07 ± 0.03 & 34.49 ± 0.10 \\
Multi-task &50\%/40\% & 98.81 ± 0.27 & 96.66 ± 0.36 & 0.20 ± 0.04 & 5.81 ± 0.01 & 14.08 ± 0.04 & 34.53 ± 0.12 \\
Multi-task & 50\%/50\% & \textbf{99.00 ± 0.11} & 96.83 ± 0.15 & \textbf{0.19 ± 0.05} & 5.79 ± 0.02 & 14.03 ± 0.05 & 34.38 ± 0.14 \\
\bottomrule
\end{tabular*}}
\label{app_multitask_inpaint}
\end{table}
\clearpage

\bibliographystyle{elsarticle-num}
\bibliography{MelodyGLM}

\end{document}